 \documentclass[final,authoryear,3p,times,twocolumn]{elsarticle}

\usepackage{graphicx}

\usepackage{amssymb}

\usepackage{rotating}
\usepackage{array}
\usepackage{color}
\usepackage{multirow}
\usepackage{float}
\usepackage{url}
\usepackage{ulem}
\usepackage{subfigure}
\usepackage{hyperref}
\hypersetup{
bookmarks=true,        
unicode=false,          
pdftoolbar=true,        
pdfmenubar=true,        
pdffitwindow=false,     
pdfstartview={FitH},    
pdftitle={N2 emission on titan},    
pdfauthor={Sonal Kumar Jain},   
pdfsubject={N2 triplet emission on Titan},   
pdfcreator={TeXlive},   
pdfproducer={TeXmakerX}, 
pdfnewwindow=true,      
colorlinks=false,        
linkcolor=blue,         
linkcolor=blue,         
citecolor=blue,        
filecolor=magenta,      
urlcolor=red            
}

\newcommand{\nt}{{$\mathrm{N_2}$}}

\newcommand{\dgr}{{$^\circ$}}

\journal{Icarus}

\begin{document}

\begin{frontmatter}


\title{Production of N$_2$ Vegard-Kaplan and Lyman-Birge-Hopfield emissions on Pluto}
\author{Sonal Kumar Jain\corref{cor1}}
\ead{SonalJain.spl@gmail.com; Sonal.Jain@lasp.colorado.edu}
\address{Laboratory for Atmosheric and Space Physics, University of Colorado, at Boulder, Colorado, USA}
\author{Anil Bhardwaj} 
\ead{Anil\_Bhardwaj@vssc.gov.in; Bhardwaj\_SPL@yahoo.com}
\address{Space Physics Laboratory,
	Vikram Sarabhai Space Centre,
	Trivandrum~695022, India}

\begin{abstract}
	We have developed a model to calculate the emission intensities of various vibrational transitions of N$_2$ triplet
	band and Lyman-Birge-Hopfield (LBH) band emissions in the dayglow of Pluto for solar minimum, moderate,
	and maximum conditions. The calculated overhead intensities of Vegard-Kaplan 
	($ A^3\Sigma_u^+ - X^1\Sigma^+_g $), First Positive ($ B^3\Pi_g - A^3\Sigma^+_u $), Second Positive 
	($ C^3\Pi_u - B^3\Pi_g $), Wu-Benesch ($W^3\Delta_u - B^3\Pi_g$), Reverse First Positive, and LBH 
	($a^1\Pi_g$-- $X^1\Sigma^+_g$) bands
	of N$_2$ are 17 (74), 14.8 (64), 2.4 (10.8), 2.9 (12.7), 2.9 (12.5), and 2.3 (10) R, respectively,
	for solar minimum (maximum) condition. We have predicted the overhead and limb intensities of 
	VK (150-190 nm) and LBH (120-190 nm) bands of N$_2$ on Pluto  for the New Horizons (NH) flyby condition that
	can be observed by Alice: the ultraviolet imaging spectrograph also know as P-Alice. The predicted 
	limb intensities of VK and  LBH bands peak at radial distance
	of $ \sim$2000 km with the value of about 5 (13) and 9.5 (22) R for solar zenith angle 60$ ^\circ $ (0$ ^\circ $), respectively.
	We have also calculated overhead and limp intensities of few prominent transition of CO Fourth Positive bands for NH flyby condition.  
\end{abstract} 

\end{frontmatter}

\section{Introduction}\label{sec:introduction}
The atmosphere of Pluto is believed to be hydrodynamically escaping and
extending to heights comparable to its radius 
\citep{Krasnopolsky99,Strobel08}. Most of the information about the atmosphere of Pluto
have come from the ground based occultation observations \citep{Elliot07,Young08}.
\cite{Schindhelm14} have provided a brief review of ultraviolet spectra observed from Pluto-Charon system by IUE.
The data from these occultation observations are used to understand Pluto's atmosphere and are instrumental
in the recent developments of general circulation models \citep{Zalucha12}. Occultation 
studies have also shown that Pluto's atmosphere has undergone a surface pressure expansion by a factor of 2
from 1988 to 2002, followed by a stabilization from  2002 to 2007 \citep{Elliot07,Young08}. The 
atmosphere of Pluto is similar to the Saturn's largest moon Titan: dominated by the N$_2$ (97-99\%),
followed by CH$_4$ (3-1\%), and a trace amount of CO. Titan's extreme ultraviolet (EUV) dayglow
spectra is dominated by N$_2$ Carroll-Yoshino (CY) Rydberg bands, while  far
ultraviolet (FUV) spectra is dominated by N$_2$ Lyman-Birge-Hopfield (LBH) and Vegard-Kaplan (VK) bands,
and N and N$^+$ lines \citep{Ajello08,Stevens11}. Pluto's dayglow spectrum is expected
to be similar to that of Titan. \cite{Summers97} have estimated the overhead emission intensities of
possible airglow features in the atmosphere of Pluto, and calculated vertical column rates of 
various emission of \nt\ and N, and N$^+$ in airglow of Pluto.

New Horizons (NH) flyby of Pluto in July 2015 and will be the first visit of a man made object to Pluto.
NH is carrying a host of instruments to understand the atmosphere and surface properties of
Pluto. One of the instruments, ALICE: an ultraviolet imaging spectrograph (Pluto-ALICE or P-ALICE), is aimed at observing Pluto
at EUV and FUV wavelength regions, and  might observe various emissions from  N$_2$ and its dissociated 
products mentioned above.
We have developed the N$_2$ triplet band emission model for atmospheres of Mars, Venus, and
Titan \citep{Jain11,Bhardwaj12b,Bhardwaj12c}. In the present study,  we have extended the
N$_2$ triplet band emission model to Pluto, and have added the calculation of 
LBH (singlet) band emissions. We have made calculations 
for solar minimum, moderate, and maximum conditions as well as for the New Horizons flyby
conditions. 
These model calculations will help in making  observations of Pluto's 
airglow by P-ALICE. In this study, we give volume emission rate and limb intensity of 
N$_2$ VK and LBH in the wavelength region 120--190 nm region, which lies in spectral range
of P-ALICE \citep{Stern08}. For the N$_2$ VK band major focus is given to transitions that lie
between 150 and 190 nm region because emissions in wavelength region 120--150 nm constitute
less than 1\% of N$ _2 $ VK emission  in wavelength region 120--190 nm 
\citep{Jain11,Bhardwaj12b,Bhardwaj12c}.  We have also reported limb and overhead intensities for CO fourth positive bands.

\section{Model}\label{sec:model}
Model atmosphere of Pluto for solar minimum, moderate, and maximum conditions, and for NH arrival condition
is taken from \cite{Strobel08}. The mixing ratios of N$_2$, CH$_4$, and CO is taken as 0.97, 0.03, and
0.00046, respectively \citep{Strobel08}. The mean Sun-Pluto distance is taken as 30 AU for solar minimum, moderate,
and maximum conditions, and 33.4 AU for the NH arrival time prediction.

We have used the solar irradiance measured at Earth (between 2.5 to 120.5 nm) by Solar EUV Experiment
(SEE, Version 10.2) \citep{Woods05,Lean11} for solar minimum (F10.7 = 68), moderate (F10.7 = 150), and 
maximum (F10.7 = 250) conditions at 1 nm spectral resolution. For the NH flyby prediction (July 2015), 
the SEE solar flux is taken as on 1 May 2011, for which F10.7 = 106; the F10.7 solar flux index value is
based on the solar cycle prediction made by \cite{Hathaway94} (\url{http://solarscience.msfc.nasa.gov/predict.shtml}).
The solar flux is scaled to the Sun-Pluto distance of 30 AU for solar minimum, moderate, and maximum conditions, 
and 33.4 AU for NH flyby predicted intensity calculations. The solar zenith angle (SZA) is taken as 
60$^\circ$ unless otherwise mentioned in the text. All distances mentioned are 
radial distances measured from the centre of the planet.

The limb intensity of various airglow emissions is calculated as 
\begin{equation}\label{eq:h}
	I = \int \left[n_l(Z) \int _{E_{th}}^{E} \left(\int_{W_{kl}}^{100} \frac{Q(Z,E) U(E,E_0)}
	{{\displaystyle\sum_{l}} n_l(Z)\sigma_{lT}(E)} \ dE_0 \right) \sigma_{il}(E) dE \right]dr,
\end{equation}
where $n_l(Z)$ is the density of the $l$th gas at altitude $ Z $, 
$\sigma_{lT}(E)$ is the total inelastic cross section for the $l$th gas,
$ \sigma_{il}(E)$ is the electron impact cross section for the $i$th state
of the $l$th gas, for which the threshold is $E_{th}$, $Q$(Z, E) is the photoelectron production rate at altitude $Z$, $W_{kl}$ is minimum excitation energy for $k$ excited state of $l$th gas, $r$ is abscissa along the horizontal line of sight, and $U(E,E_0)$ is the two-dimensional Analitical Yield Spectra (AYS),
which embodies the non-spatial information of degradation process. 
It represents the equilibrium number of photoelectrons per unit energy at an
energy $E$ resulting from the local energy degradation of an incident
electron of energy $E_0$ \citep{Bhardwaj90b,Bhardwaj96,Bhardwaj03,Bhardwaj99b,Bhardwaj09}.
While calculating the line of sight intensities
of various LBH transition, we have taken absorption by N$_2$ and CH$_4$ into consideration, because below 140 nm, absorption by atmospheric gases can significantly affect the calculated limb intensities.

Electron impact excitation following XUV photoionization is the major source of \nt\ triplet band emissions. 
The details of the model calculation of N$_2$ triplet band emissions is provided in our earlier studies
on Mars \citep{Jain11}, Venus \citep{Bhardwaj12b}, and Titan \citep{Bhardwaj12c}. In the present paper
we have extended  the model to include the N$_2$ LBH band emission calculation. \cite{Johnson05} have
reported the excitation cross section of a$^1\Pi_g $ state.
\cite{Young10} have measured emission cross section for the a$^1 \Pi_g$ ($\nu'= 3$) -- X$^1\Sigma_g^+ $ ($\nu''= 0$)
and a$^1 \Pi_g$ ($\nu'= 2$) -- X$^1\Sigma_g^+ $ ($\nu''= 0$) transitions. The shape and magnitude of their relative
emission cross section of a(3,0) and a(2,0) emission is consistent with the results of \cite{Johnson05}. \cite{Young10}
have not given emission cross section for entire LBH band system, though they suggested the emission cross section 
value of $ 6.3 \pm 1.1 \times 10^{-16} $ cm$ ^2 $ at 100 eV.
The electron impact excitation cross section of singlet (a$^1 $, a$'^{1}$, w$ ^1 $) states  are taken 
from \cite{Johnson05} and 
fitted analytically for ease of usage in the model  \citep[see][and references therein]{Jain11}.
We have included radiative cascading between three lowest singlet states of N$_2$ (a$ ^1\Pi_g $, a$ '\Sigma_u^- $, 
and w$ ^1\Delta_u $). For $a$ state, vibrational levels which lie above $ \nu = 6 $ are not considered since
these vibrational levels predissociate, and we have excluded the cascade contributions from $ a' $ and $ w $
states vibrational levels which lie above $ \nu = 6 $ of the $ a $ state \citep{Eastes00,Eastes11}. The 
radiative transition data for the $ a' \leftrightarrow a$, $ w \leftrightarrow a$, and $ a \rightarrow X $ are
taken from  \cite{Gilmore92}. For the $ a' \rightarrow X  $ transition, lifetime of 17 ms is taken for all
vibrational level \citep{Eastes00}. The collisional quenching rates of $ a, a', w $ states are taken as
described by \cite{Eastes00}.

\section{Results}\label{sec:result}
\begin{figure}[h]
	\centering\includegraphics[width=20pc]{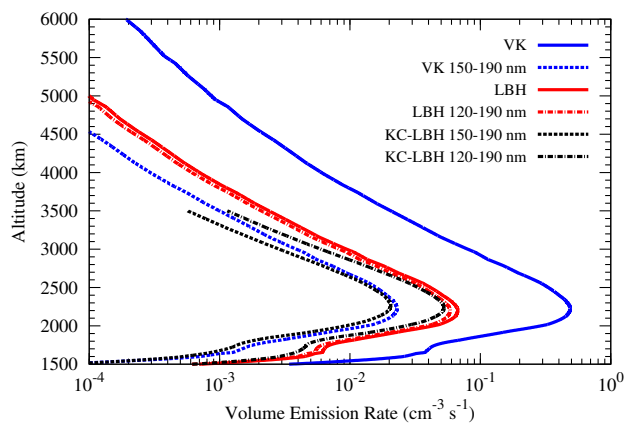}
	\caption{Calculated volume emissions rates of N$_2$ VK (total), LBH (total),  N$_2$ VK (150--190 nm) and LBH 
		(120--190 nm) bands for solar moderate condition at SZA = 60$^\circ$ and Pluto-Sun distance of 30 AU. 
		Black curves show the emission rate calculated using the M1 model atmosphere of \cite{Krasnopolsky99} for 
		VK (150--190 nm) and LBH (120--190 nm) bands.}
	\label{fig:ver}
\end{figure}

The volume emission rates of vibrational transitions of various triplet and LBH bands 
of N$_2$ on Pluto are calculated for the three solar conditions.
Figure~\ref{fig:ver} shows the volume emission rates of N$_2$ VK (150-190 nm) and LBH 
(120-190 nm) bands for solar moderate condition at SZA = 60$ ^\circ $. Total emission
rates of N$ _2 $ VK and LBH bands are also shown in the figure.
\begin{table*}[t]
	\caption{Height-integrated overhead intensities of N$_2$ triplet and LBH bands emissions on Pluto
		for solar minimum, moderate, and maximum conditions at SZA = $ 60^\circ $ and 30 AU, and the prediction 
		for New Horizons flyby condition.}
	\small
	\begin{tabular*}{\textwidth}{@{\extracolsep{\fill}}lllll}
		\hline
		\multirow{3}{3cm}{Band} & \multicolumn{4}{c}{Intensity (R)} \\
		\cline{2-5}
		&  Minimum & Moderate  & Maximum	&  NH\footnotemark[1] \\
		&  	& 	 & &Prediction	\\
		\hline 
		\multicolumn{5}{c}{\it Triplet bands} \\ \noalign{\smallskip}
		VK ($A \rightarrow X$)  (137--1155 nm) & 17   & 37 (30.3)\footnotemark[2]  & 74  & 25.3  \\
		\hspace{1cm}137--190 nm & 0.8 & 1.7 (1.4)  & 3.5  & 1.2   \\
		\hspace{1cm}200--300 nm & 5.9  	& 13 (10.5) 	& 25.8    	& 8.8   \\
		\hspace{1cm}300--400 nm & 6.6 	& 14 (11.8) 	& 28.9  	& 9.8   \\
		\hspace{1cm}400--800 nm & 3.7   & 8 (6.6)  	& 16    	& 5.5   \\
		1P ($B \rightarrow A$) (263--94129 nm)	& 14.8  	& 32 (26.2) & 64.4 & 21.9   \\
		2P ($C \rightarrow B$) (268--1140 nm)	& 2.4   & 5 (4.4) & 10.8  & 3.7   \\
		$W \rightarrow B$ (399--154631 nm)	& 2.9  	& 6 (5.1) & 12.7   & 4.3  \\
		$B' \rightarrow B$ (312--37699 nm) 	& 1.2  	& 2.7 (2.2) & 5.4    & 1.8   \\
		R1P  ($A \rightarrow B$) (739--74175 nm)	& 2.9   & 6.3 (5.1)   & 12.5   & 4.2  \\
		$E \rightarrow A$ (207--303 nm)		& 3.3E-2\footnotemark[3]  & 7.2E-2 (5.9E-2) & 0.14   	& 4.9E-2 \\
		$E \rightarrow B$ (259--483 nm)		& 5.2E-3    & 1.3E-2 (9.3E-3)  & 2.3E-2    & 7.8E-3  \\
		$E \rightarrow C$ (1113--10127 nm)	& 1.9E-2 & 4.2E-2 (3.5E-2) & 4.4E-2 & 2.9E-2  \\[10pt]
		\multicolumn{5}{c}{\it Singlet bands} \\ \noalign{\smallskip} 
		LBH ($a \rightarrow X$)  (120--260 nm) & 2.3   & 4.9 (4)  & 10   & 3.4  \\
		\hspace{1cm}120--190 nm & 2  & 4.3 (3.5)   & 8.8  & 3  \\
		\hline\noalign{\smallskip} 
	\end{tabular*}
	VK = Vegard-Kaplan; 1P = First Positive; 2P = Second Positive, LBH = Lyman-Birge-Hopfield band, 
	R1P = Reverse First Positive \\
	\footnotemark[1]{Prediction for New Horizons arrival on Pluto (F10.7 = 106) for Pluto-Sun distance
		of 33.4 AU.}\\
	\footnotemark[2]{Values in the parenthesis are overhead intensities calculated by using
		the M1 model atmosphere of \cite{Krasnopolsky99} for solar moderate condition.}\\
	\footnotemark[3]{3.3E-2 =  $ 3.3 \times 10^{-2}$.}
	\label{tab:oi-trip-sing}
\end{table*}
Emission rates of both VK (150--190 nm) and LBH (120--190 nm) bands peak at radial distance 
of about 2200 km, with a value
of 0.02 and 0.06 cm$^{-3}$ s$^{-1}$, respectively. Table~\ref{tab:oi-trip-sing} shows the total 
height-integrated overhead intensity for Vegard-Kaplan (VK) ($A \rightarrow X$), First positive 
($B \rightarrow A$), Second Positive ($C \rightarrow B$),  Herman-Kaplan  (HK) ($E \rightarrow A$),
$E \rightarrow B$, Reverse First Positive ($A \rightarrow B$), 
$E \rightarrow C$, and  Lyman-Birge-Hopfield ($ a \rightarrow X $) bands of N$_2$ at 
SZA=60$ ^\circ$, for minimum, moderate, and maximum solar conditions. Since many
bands of N$_2$ span a large wavelength region, the overhead intensities of VK
and LBH bands in different wavelength regions are also given in the Table~\ref{tab:oi-trip-sing}.
\cite{Summers97} have calculated volume emission rates of various emissions of N$_2$ and N on
Pluto for moderate solar activity condition at Sun-Pluto distance of 30 AU. Their calculated
volume emission rates peaks at $\sim$2000 km radial distance. For N$_2$ LBH bands,
\cite{Summers97} have calculated an overhead intensity of about 5.7 R for CH$_4$ mixing ratio
of $5 \times 10^{-4}$. Our calculated overhead intensity of N$_2$ LBH is 4.9 R for solar moderate condition.  The minor difference in the two studies may largely be due to the  cross section for N$_2$ LBH band used in the two studies and the input solar flux model.

\begin{figure}[h]
	\centering\includegraphics[width=20pc]{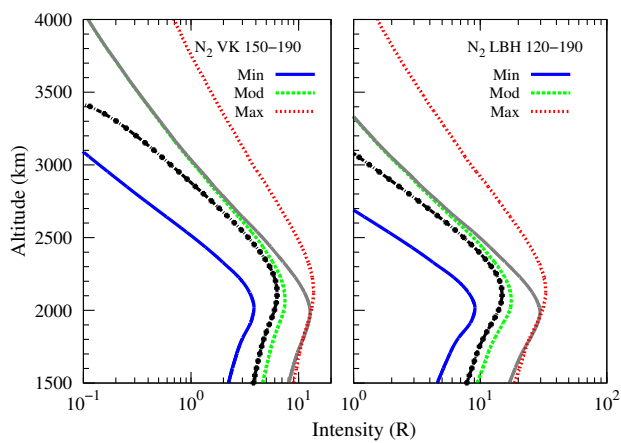}
	\caption{Calculated limb intensities of N$_2$ VK (150--190 nm) and LBH (120--190 nm) bands for solar minimum, moderate,
		and maximum conditions for SZA = 60$^\circ$ and Pluto-Sun distance of 30 AU.  Solid gray curve shows the intensity calculated at SZA = 0$ ^\circ $ for solar moderate
		condition. Black curve with symbols shows the
		intensity calculated using M1 model atmosphere from \cite{Krasnopolsky99} for solar moderate condition at
		SZA = 60$ ^\circ $.}
	\label{fig:limbint}
\end{figure}
Figure~\ref{fig:limbint} shows the calculated limb intensities of N$_2$ VK (150-190 nm) and LBH (120-150 nm)
bands on Pluto for solar minimum, moderate, and maximum conditions. The limb intensity of N$_2$ VK (LBH) band  
peaks at radial altitude of about 2025, 2075, and 2125 km with magnitude of 3.8 (9.1) R, 7.4 (17.5) R, and 
13.8 (32.9) R for solar minimum, moderate, and maximum conditions, respectively. Calculated
limb intensities of N$_2$ VK and LBH bands at SZA  = 0$^\circ$ are also shown for solar moderate
condition  in  Figure~\ref{fig:limbint}, which peak at 1995 and 1988 km with values of 12.8 and 29.7 R, respectively.
We have also calculated limb intensities for other solar zenith angles. For SZA = 30$^\circ$ (45$^\circ$) the calculated limb intensities of \nt\ VK 
(150--190 nm) and LBH (120--190 nm) bands peak at radial altitude of $\sim$2000 (2020) km with values of 11.3 (9.6) R and 26.5 (22.6) R, 
respectively.

\section{Discussion}
\subsection{Effect of model atmosphere}
\cite{Krasnopolsky99} have developed a detailed photochemical model of Pluto's atmosphere. Based
on the occultation curve data, they have used two atmospheric models in their calculation for
solar moderate condition and SZA = 60$^\circ$;
one assuming atmospheric tropopause at 1195 km (M1), and other assuming planet's surface at 1195 km (M2).
In both models N$_2$ density is similar up to about 2000 km, and above  this
distance N$_2$ is  higher in model M1. To evaluate the effect of model atmosphere on 
the calculated intensity, we have taken model atmosphere M1 from the study of \cite{Krasnopolsky99}
for the moderate solar activity condition. Figure~\ref{fig:ver} shows volume emission rates
of N$_2$ VK (150-190 nm) and LBH (120-190 nm) bands calculated by using the M1 model atmosphere from
\cite{Krasnopolsky99} in addition to those using model atmosphere of \cite{Strobel08}.
The emission rate peaks at about same radial distance for the two model atmospheres,
but the magnitude of emission rate at the peak is about 15\% higher on using the model atmosphere of 
\cite{Strobel08}. 
Table~\ref{tab:oi-trip-sing} presents the overhead intensities calculated by using the M1 model 
atmosphere from \cite{Krasnopolsky99}.

\subsection{Prediction}
P-ALICE instrument is an imaging spectrograph aboard the New Horizons mission to Pluto/Charon
and the Kuiper belt, with a spectral passband of 52--187 nm \citep[see][for details]{Stern08}.
The main objective of P-ALICE instrument is to perform spectroscopic investigations of Pluto's
atmosphere and surface at EUV and FUV wavelengths. The spectral passband of
P-ALICE is very similar to the Cassini-UVIS instrument, which observed both N$_2$ VK (150--190 nm) 
and LBH  (120--190 nm) bands in the dayglow of Titan \citep{Stevens11}. New Horizons 
flyby of Pluto will occur around July 2015. For the conditions similar (see Section~\ref{sec:model})
to NH flyby of Pluto, we predict the intensity of N$_2$ VK and LBH bands within the spectral
passband of P-ALICE. With given baseline trajectory \citep[cf.][]{Guo05}, for a limb observation solar 
zenith angle variation may not be significant, hence we have predicted limb intensities for 
fixed solar zenith angles of 0$^\circ$, and 60$^\circ$.

As discussed earlier, intensity of N$_2$ VK bands in 120-150 nm  wavelength region is very small (0.02\%), 
which make them difficult to observe in the dayglow spectra, 
and explain their absence in the dayglow spectrum of Titan \citep{Stevens11,Bhardwaj12c}.
However,  N$_2$ VK (150--190 nm) and LBH (120--190 nm) bands  are observed on Titan \citep{Stevens11}. 
For the NH flyby condition, the calculated overhead intensities of various triplet and LBH bands of N$_2$ is given in
Table~\ref{tab:oi-trip-sing}. The overhead intensities of N$_2$ VK (150--190 nm) and LBH (120--190 nm) bands are
1.2  and 3 R, respectively.
Prominent emissions of N$_2$ VK and LBH bands in wavelength region 150--190 and 120--190 nm
are given in Table~\ref{tab:n2-oi}. The N$_2$ VK (7-0), (8-0), and (9-0) are dominant emissions
in wavelength region 150--190 nm with intensities of 0.21, 0.2, and  0.11R, respectively.
In 120--150 nm region, N$_2$ LBH (2-0), (3-0), (4-0), and (5-0) bands are the dominant emissions
with intensities of 0.18, 0.22, 0.18, and 0.11 R, respectively. On Titan, our calculations suggest
that intensities of N$_2$ VK (8-0) at 165.5 nm and (11-0) at 156.3 nm is about 2 to 3 orders of
magnitude higher than that of CI 165.7 and 156.1 nm lines \citep{Bhardwaj12c}, which were 
misidentified as possible source for carbon emissions near 165 and 156 nm
\citep[cf.][]{Ajello08,Stevens11}. On Pluto, 
we expect N$_2$ VK (8-0) and (11-0) bands intensities to be higher than that of CI 165.7
and 156.1 nm lines. 

\begin{figure}[h]
	\centering\includegraphics[width=20pc]{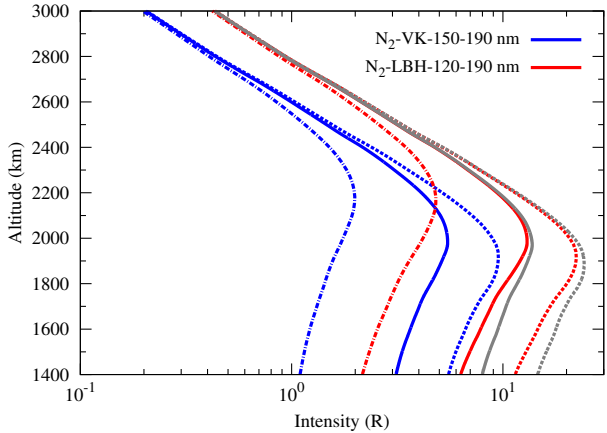}
	\caption{Predicted limb intensities of N$_2$ VK (150--190 nm) and LBH (120--190 nm) bands, for conditions
		similar to New Horizons flyby of Pluto in July 2015 (F10.7 = 106) at 33.4 AU (see text for detail). Solid 
		curves are for SZA = 60$ ^\circ $, dashed curves are for SZA = 0$ ^\circ $, and dashed dotted curves are for SZA = 90$ ^\circ $.  Gray solid and dashed
		curves show the calculated intensity of LBH (120--190 nm) band for optically thin ($ \tau = 0 $) condition at solar zenith angles 
		60$ ^\circ $ and 0$ ^\circ $, respectively.}
	\label{fig:limbint-prediction}
\end{figure}
We have also calculated overhead intensity and limb intensities of CO Fourth Positive (A$^1\Pi$ -- X$^1\Sigma^+$) band by electron impact excitation of CO and by fluorescent scattaring. The electron impact emission cross section are taken from \cite{Avakyan98} and fluorescence efficiencies for low solar activity conditions are taken from \cite{Tozzi98}. The overhead intensities of CO Fourth Positive (0,1; 159.7 nm), (1,4; 172.9 nm), (2,2; 157.6 nm), (3,0; 144.7 nm), and (4,0; 141.9 nm) emissions due to electron impact excitation (fluorescent scattering) are  1.4E-4 (2), 1.1E-4 (1.4), 1.2E-4 (1.1), 3.6E-4 (2.5), and 1.5E-4 (0.8) R, respectively, for NH flyby condition (see Table~\ref{tab:n2-oi}).
%

\begin{table}
	\centering
	\caption{Predicted height-integrated overhead intensities 
		of N$_2$ VK, LBH, and CO fourth positive bands on Pluto for New Horizons flyby condition at SZA = 60$ ^\circ $. Prominent transitions
		between 120 and 190 nm, which lie in the spectral band of P-ALICE, only are given.}
	\tiny
	\begin{tabular}{ccccccc}
		\hline \noalign{\smallskip} 
		Band & Band Origin & \multicolumn{1}{c}{Intensity} & & Band & Band Origin & \multicolumn{1}{c}{Intensity} \\ \noalign{\smallskip} 
		($ \nu' - \nu''$) & nm &  (R) &&($ \nu' - \nu''$) & nm &  (R)  \\ \noalign{\smallskip} 
		\cline{1-3} \cline{5-7}
		\noalign{\smallskip}
		\multicolumn{3}{c}{Vegard-Kaplan ($A^3\Sigma^+_u$-- $X^1\Sigma^+_g$)} && \multicolumn{3}{c}{Lyman-Birge-Hopfield ($a^1\Pi_g$-- $X^1\Sigma^+_g$)}\\[5pt]
		3-0	&	185.3 	& 	2.1E-2	 &&  3-9	&	185.4	&	7.9E-2   \\
		4-0	&	180.8  	& 	4.7E-2 	 &&  4-0	&	132.5	&	1.8E-1   \\
		5-0 	&	176.5 	&	8.2E-2   &&  4-2	&	141.2	&	8.9E-2   \\
		5-1	&	184.1	&	4.3E-2   &&  4-4	&	150.8	&	5.9E-2   \\
		6-0	&	172.6 	&	1.3E-1   &&  4-5	&	156.0	&	2.5E-2   \\ 
		6-1	&	179.8 	&	4.5E-2 	 &&  4-6	&	161.6	&	1.1E-2   \\
		7-0	&	168.9	&	2.1E-1	 &&  4-8	&	173.6	&	1.7E-2   \\
		7-1	&	175.8	&	3.7E-2	 &&  4-9	&	180.1	&	5E-3     \\
		8-0	&	165.5	&	2E-1     &&  5-0	&	129.9	&	1.1E-1   \\
		9-0	&	162.2	&	1.1E-1   &&  5-1	&	133.9   &	3.7E-2   \\
		10-0	&	159.2	&	5.1E-2 	 &&  5-2	&	138.2	&	3.5E-2   \\
		11-0	&	156.3	&	2.2E-2 	 &&  5-3	&	142.7	&	2.3E-2   \\ 
		12-0	&	153.6	&	1.1E-2 	 &&  5-4	&	147.4	&	3.5E-2   \\  [3pt]
		\multicolumn{3}{c}{Lyman-Birge-Hopfield ($a^1\Pi_g$-- $X^1\Sigma^+_g$)} && 4-7	&1674	&4E-2  \\[3pt]
		0-1	&	150.1	&	3.3E-2	 &&  5-6	&	157.6	&	3.9E-2   \\
		0-2	&	155.5	&	4.9E-2   &&  5-8	&	169.0	&	2E-2     \\
		0-3	&	161.2	&	4.4E-2   &&  5-9	&	175.2	&	3.1E-2   \\
		0-4	&	167.2	&	2.7E-2   &&  6-0	&	127.3	&	5.4E-2   \\
		1-0	&	141.6	&	8.0E-2   &&  6-1	&	131.2	&	5E-2     \\
		1-1	&	146.4	&	1.2E-1   &&  6-3	&	139.6	&	3.6E-2   \\
		1-2	&	151.5	&	4.6E-2   &&  6-5	&	148.9	&	2.6E-2   \\
		1-4	&	162.7	&	4E-2	 &&  6-7	&	159.2	&	1.4E-2   \\
		1-5	&	168.8	&	7.6E-2   &&  6-8	&	164.8	&	1.6E-2   \\
		1-6	&	175.2	&	6.2E-2   && 6-10	&	176.9	&	1.9E-2   \\
		1-7	&	182.1	&	3.3E-2   && 6-11	&	183.5	&	1E-2     \\
		2-0	&	138.4	&	1.8E-1	 &&  6-3	&	139.6	&	3.6E-2   \\
		2-1	&	143.0	&	9.4E-2	 &&  6-5	&	148.9	&	2.6E-2   \\ 
		2-3	&	153.0	&	8.5E-2	 &&  6-7	&	159.2	&	1.4E-2   \\ 
		2-4	&	158.5	&	6.1E-2	 &&  6-8	&	164.8	&	1.6E-2   \\ 
		2-6	&	170.3	&	3.8E-2	 && 6-10	&	176.9	&	1.9E-2   \\ 
		2-7	&	176.8	&	8.6E-2   && 6-11	&	183.5	&	1E-2     \\ 
		2-8	&	183.8	&	7.4E-2   &&  \multicolumn{3}{c}{CO Fourth Positive ($A^1\Pi$-- $X^1\Sigma^+$)}  \\ 
		3-0	&	135.4	&	2.2E-1	 &&  0-1	& 	159.7	&	2	 \\
		3-1	&	139.8	&	1.4E-2   &&  1-4	&	172.9	&	1.42	 \\
		3-2	&	144.4	&	5.5E-2   &&  2-2	&	157.6	&	1.1	 \\
		3-3	&	149.3	&	6.3E-2   &&  3-0	&	144.7	&	2.5	 \\
		3-5	&	160.0	&	7.1E-2   &&  4-0	&	141.9	&	0.83	 \\
		3-6	&	165.8	&	4.3E-2   &&  5-0	&	139.2	&	0.79	 \\	
		3-8	&	178.5	&	4.4E-2   &&  5-1	&	143.5	&	1.36	 \\	
		\hline
	\end{tabular}
	\label{tab:n2-oi}
\end{table}

Figure~\ref{fig:limbint-prediction} shows the limb intensity of N$_2$ VK (150--190 nm)
and LBH (120--190 nm) bands predicted for NH flyby condition.
Calculated limb intensities of VK and LBH bands peak at the radial distance of 2000 km with
values of $ \sim$5 and 10 R, respectively. For SZA = 0$^\circ$, the calculated intensity
increases by a factor of 2 and the altitude of peak emission decreases by about 75 km and peaks at
about 1920 km.

\begin{figure}[h]
	\centering\includegraphics[width=20pc]{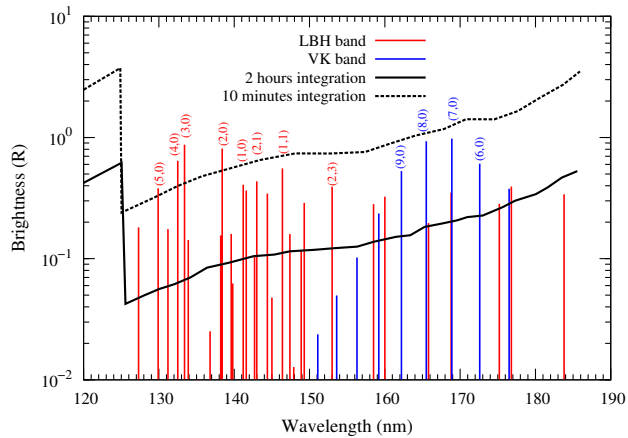}
	\caption{Predicted limb intensities (at SZA = 60$^\circ$) of various transitions of N$_2$ LBH (red), 
		VK (blue), CO Fourth Positive (green) bands at 2000 km radial distance. Prominent emission lines of LBH, VK, and CO fourth positive  bands are marked with vibrational transitions at top of them while symbol shows the predicted overhead intensities of the same transition (see Table~\ref{tab:n2-oi}).
		The expected P-ALICE SNR $\sim$ 10 for
		10-minutes (gray) and 2-hours (black) integrations at a range of 10$^5$ km is also shown in figure
		\citep[taken from Figure 1 of][]{Stern08}.} 
	\label{fig:snr}
\end{figure}
For P-ALICE to observe the \nt\ LBH and VK bands in the dayglow of Pluto, it is very essential that
emission intensities must be higher than the sensitivity of the instrument, and also for a given observation,
the signal to noise ratio (SNR) should be high enough to detect any meaningful data.
In Figure~\ref{fig:snr}, we present the predicted limb intensities of prominent transitions of \nt\ LBH and VK, and CO fourth positive bands at 2000 km, along with expected P-ALICE SNR $\sim$ 10 for 10-minutes and 2-hours 
integrations \citep[taken from Figure 1 of][]{Stern08}. The overhead intensities of \nt\ LBH and VK, and CO Fourth positive bands are also shown in Figure~\ref{fig:snr}, which shows that overhead intensity of CO fourth positive is higher than both \nt\ LBH and VK bands.
The sensitivity analysis  given by \cite{Stern08} is not related to viewing geometry and their estimated brightness features are for unresolved disk and limb. 
We found that for 2-hr integration, P-ALICE could detect a number of prominent transitions of \nt\ LBH and VK bands, however for low integration 
(10-minutes) only a very few transitions are above SNR level of 10.  
This suggests that integration time for P-ALICE observations should be higher for it to detect
bright emissions of \nt\ LBH and VK bands.

\subsection{Comparison with Titan's \nt\ emissions}
\begin{figure}[h]
	\centering\includegraphics[width=20pc]{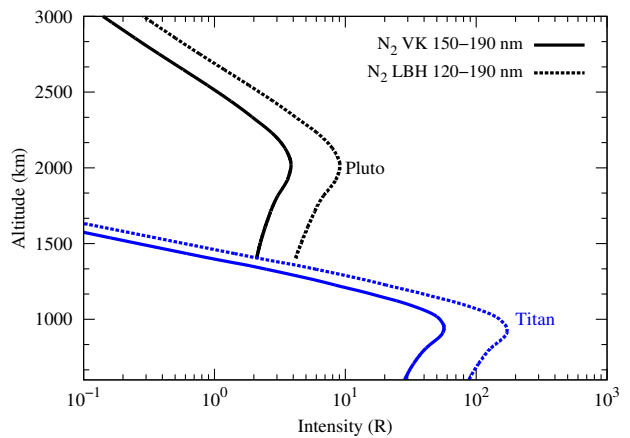}
	\caption{Limb intensities of \nt\ VK (120--190 nm) and LBH (150--190 nm) bands on Titan (blue curves) and Pluto (black curves) for solar minimum conditions and for SZA = 60\dgr. The Sun-Titan and Sun-Pluto distances are taken as 9.7 and 30 AU, respectively.}
	\label{fig:limb-compare}
\end{figure}

UVIS onboard Cassini spacecraft have observed \nt\ VK and LBH bands emissions on Titan \citep{Stevens11}. As mentioned earlier, we have developed model for \nt\ emissions on Titan \citep{Bhardwaj12c}. 
Figure~\ref{fig:limb-compare} shows the calculated limb intensities of \nt\ VK (150-190 nm) and LBH (120-190 nm) bands in the atmospheres of Titan and Pluto for solar minimum condition (F10.7=68) and SZA = 60\dgr. Figure~\ref{fig:limb-compare} clearly depicts that \nt\ emissions intensities on Titan is an order of magnitude higher than that on Pluto, reflecting closer proximity of Titan to Sun. The altitude of peak emissions on the Titan and Pluto differs by about 1000 km, which mostly due to the extended atmosphere of Pluto.

\section{Conclusions}
We present the calculation of N$ _2 $ dayglow emissions in the atmosphere of Pluto.
Our calculations predicted peak limb intensity of 5 (10) R for N$_2$ VK 150--190 (LBH 120--190) nm for the NH
flyby conditions. However, the predicted intensities depend upon various input model parameters and the 
observed intensities would depend on conditions at the time of measurements. Based on our 
current understanding of Pluto's atmosphere for solar minimum and maximum conditions, variability
for the peak limb intensity in the range 4 to 13.7 R (9 to 33 R) for N$ _2 $ VK (LBH) band is expected. Our calculation shows that overhead intensity of CO 4P bands for NH flyby condition is higher than that of \nt\ VK and LBH bands. 
We hope that the NH mission will significantly improve our understanding of  Pluto's atmosphere,
and suggest conducting the limb observations of P-ALICE about 2000 km radial distance where airglow emissions peak and with enough integration time to detect prominent transitions of \nt\ LBH and VK bands.
%

\section*{Acknowledgement}
\vspace{-0.3cm}
{\small The work of SKJ was supported by DST INSPIRE Faculty program at SPL, VSSC, while work of AB was supported by ISRO.}



%



\newpage

\renewcommand{\thefootnote}{\fnsymbol{footnote}}

\clearpage

\end{document}